\documentstyle[11pt,aspconf]{article}
\begin{document}

\title{Associated Absorption at Low and High Redshift}
\author{Martin Elvis$^1$, Smita Mathur$^1$, Belinda J. Wilkes$^1$,
Fabrizio Fiore$^{1,2}$, Paolo Giommi$^2$ \& Paolo Padovani$^2$}
\affil{1: Harvard-Smithsonian Center for Astrophysics, 60 Garden St.,
Cambridge, MA 02138, USA\\ 2: SAX Data Center, Roma, Italy}

\begin{abstract}
Combining information on absorbing material in AGN from X-ray and the
UV creates a powerful investigative tool. Here we give examples from
both low and high redshift.
\end{abstract}

\keywords{Absorption, X-ray, Optical}

\section{Introduction}

At low redshift we have found that the ionized (``warm'') X-ray
absorbers and the associated UV absorbers in two radio-loud quasars
were due to the same material : an X-ray quiet quasar 3C351 ( Mathur
et al. 1994) and a red quasar 3C212 (Elvis et al., 1994, Mathur 1994).
In both cases the absorber is situated outside the broad emission line
region (BELR), is outflowing, and is highly ionized. This delineates a
new nuclear component in lobe-dominated, radio-loud quasars. Could the
same component explain all the X-ray and UV absorption in AGN seen
over the past 20 years and more (Anderson 1974, Ulrich 1988)?

We have recently tested this generalization using the best studied of
all AGN, NGC5548. We applied the same photoionization modeling method
(Mathur et al., 1994) to the X-ray and UV absorbers in NGC5548 to
determine whether consistent values for the abundances of all the
observed ions could be obtained. In NGC5548 the model must meet two
extra requirements: it must not lead to a density for the absorber in
conflict with its recombination time; and the distance of the absorber
from the continuum source must not conflict with the well-determined
BELR size.

At high redshifts X-ray absorption and rest frame UV absorption have
been found together in a number of radio-loud quasars. The low energy
X-ray cut-offs in these objects are likely to be due to their
environment.  The absorption seen in the high-z quasars may be similar
to the low-z `X/UV' absorption, but on a larger scale.

\section{Testing the X/UV models with NGC5548}

ASCA observations confirm the presence of an ionized absorber in
NGC5548 with equivalent N$_{H}= 3.8 \times 10^{21} cm^{-2}$ (Fabian et
al. 1994a), and resolving the OVII and OVIII absorption edges. An Fe-K
edge is not detected ($\tau_{Fe-K} \leq 0.1$). HST finds blueshifted
UV absorption lines (Korista et al. 1995). The CIV and N~V doublets,
and an associated Ly$\alpha$ absorption line ($N_{HI} \geq 4\times
10^{13}$ cm$^{-2}$) are all clearly seen in the mean FOS spectrum.

We searched (using CLOUDY, Ferland 1991) for a photoionized absorber
satisfying both X-ray and UV constraints. Figure 1 shows the
ionization fractions of OVII and OVIII as a function of ionization
parameter, $U$. We used the de-reddened continuum for NGC5548 and
assumed solar abundances (Grevesse \& Andres 1989) and density
$n=10^7$ atoms cm$^{-3}$. The ASCA constraints on the fractional
ionization of OVII and OVIII (Fig.1, thick lines) allow only a narrow
range of $U$, $2.2<U<2.8$.

\begin{figure}
\vspace*{-0.98in}
\centerline{
\epsscale{.75}
\plotone{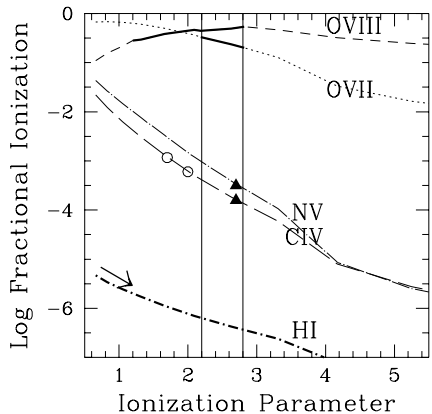}
}
\vspace*{-0.8in}
\caption{Ionization fractions of OVII, OVIII, CIV, NV and
HI as a function of U. The thick lines mark the observed
ranges for OVII and OVIII (ASCA). Triangle: HST values for CIV and NV;
$\circ$: IUE range. The HST range for HI is large,
represented by the thick curve. Arrow: HUT upper limit.  The vertical
lines define the best fit model parameter: 2.2 $<$ U $<$ 2.8.}
\end{figure}

In the mean HST spectrum, the CIV doublet ratio is 3.8$\pm 0.2$,
putting them off the linear portion of the curve of growth.
 A consistent solution for all
three ions, CIV, NV and HI is obtained for b=40~km~s$^{-1}$, with only
a small tolerance for both UV and X-ray constraints to be met (see
Mathur et al. 1995 for the details of the model). The matching of the
five ion abundances leads us to conclude that the UV and X-ray
absorbers in NGC5548 are one and the same.

An additional test of the model is now available.  The HST Ly-$\alpha$
HI column density is highly uncertain: $13<log N_{HI}<18$, while the
model values are tightly constrained, from $15.2$ to $15.4$. Mathur et
al. (1995) noted that a Lyman edge absorption would be observed if
$log N_{HI}>16.3$, and would be detectable by HUT. In the event HUT
did not find a Lyman edge (Kriss et al. 1996), implying
N$_{HI}<10^{\sim16.3}$, close to our best fit value. This strengthens
our X/UV model. OVI absorption would provide another strong
test. Unfortunately the HUT spectrum seems to have low s/n in OVI,
although the OVI absorption doublets may be present.

Our model is also consistent with the ASCA limit on an Fe-K X-ray
absorption edge of $\tau < 0.1$, implying N$_{FeXVII}< 2\times
10^{18}$ cm$^{-2}$ (for solar abundance). For our best fit model the
dominant stage of iron is FeXVII. (This is common. FeXVII dominates
over a wider range of $U$ than other ionization states since it is
neon-like and so more stable than other iron ions.) We find $\log
f_{FeXVII}= -0.77$, implying N$_{FeXVII}= 3\times 10^{16}$, far below
the ASCA limit.

The absence of an Fe-K absorption edge affects another model. The warm
gas above and below the torus that electron scatters and polarizes
light from the BELR into our line of sight in many Seyfert 2 galaxies
is a natural candidate for the ionized X-ray absorbers (Krolik \&
Kriss, 1996). In unified schemes this gas will be seen pole-on in
Seyfert 1 galaxies and will cause absorption. Krolik \& Kriss (1996)
predict an Fe-K or an Fe-L edge of optical depth $\geq 0.1$. The
absence of these features in the NGC5548 ASCA spectrum pushes these
models to higher $U$ and so lower $n_e$ and larger size. Our X/UV
absorber modeling finds smaller column density material at a lower
ionization state, and so is due to some other nuclear component.

Netzer (1996) has modeled X-ray absorbers in a similar way to Mathur
et al. (1995) but predicts that the UV lines will show N(NV)$>$N(CIV),
in contradiction to the observations of NGC5548, and concludes that two
separate absorbers are needed in NGC5548. However, Netzer uses a
continuum with very steep EUV slope. The  observed continuum of
NGC5548 instead gives N(NV)$<$N(CIV) (Mathur et al., 1995), as observed. This
illustrates the danger of comparing results using differing
assumptions.

Our model, together with the reverberation mapping variability
constraints, leads us to understand the physical properties of the
absorber. The absorber is highly ionized,
has high column density,
low density,
and is situated outside the CIV emitting region
 The gas is outflowing with a mean velocity of 1200$\pm
200~km~s^{-1}$ (relative to the host, Heckman 1978), and has a
corresponding kinetic luminosity of $\sim 10^{43} erg s^{-1}$. A
scenario in which the absorbing material comes off a disk, and is
accelerated by the radiation pressure of the continuum source may
explain the observed properties of the absorber.

We can now generalize our unification of UV and X-ray absorbing
outflows from the lobe dominated radio-loud quasars to include
radio-quiet Seyfert galaxies. This may also provide a link to the
radio-quiet BALQSOs, which show unexpectedly strong X-ray absorption
(Mathur, Elvis \& Singh 1996, Green \& Mathur 1996). This analogy
suggests that the X-ray/UV absorbers in radio-quiet AGN may be viewed
close to edge-on, which would be a valuable known parameter if it can
be independently supported.

\section{Absorption in High-Z Quasars}

A few ROSAT PSPC spectra of high redshift (z$\sim$3) quasars showed
strong low energy cut-offs, suggesting strong obscuration (Elvis et
al., 1994). A search of the whole PSPC pointed archive (Fiore et al.,
1996) has now shown that only radio-loud quasars have X-ray colors
 suggesting cut-offs; so {\em low energy X-ray cut-offs are
associated with the quasars}, and not with intervening systems (since
those would affect radio-quiet and radio-loud equally). Moreover, among
radio-loud quasars those at high redshift are more cut-off than those
at low z; so {\em the X-ray cut-offs show evolution with
cosmic epoch.}

Investigating the optical and radio properties of the 11 quasars with
ROSAT cut-offs (Elvis et al., 1996) we find that {\em all} have
associated absorption lines in their optical/ultraviolet spectra
and/or show reddening associated with the quasar. We conclude that
{\em absorption is highly likely to be the cause of the X-ray
cut-offs} too. The implied X-ray column densities are a few$\times
10^{22}$~cm$^{-2}$.

Moreover, the higher redshift quasars are Gigahertz Peaked Spectrum
source candidates suggesting that the absorbing material is extended
on the scale of the radio sources (i.e. pc - kpc).

There are several trends within the sample: going from low to high
redshift and luminosity we find a related change from low to high
ionization, and from low to high compactness (as indicated by radio
size and cut-off frequency). Interestingly, the ionization parameter
and column densities are similar to those expected from a large
`cooling flow' ionized by a quasar. Even these pressures are
insufficient to thermally confine the radio sources, but ram pressure
can slow down their expansion. The suggestive picture that emerges is
of radio sources that are both young and frustrated (Fanti 1990) by a
high pressure surrounding medium.

\acknowledgments
BJW gratefully acknowledges the financial support of NASA contract
NAS8-39073 (ASC) and SM of NASA grant NAGW-4490 (LTSA), and ME of NASA
grant NAG5-1356 (ROSAT).

\end{document}